\documentclass[useAMS,usenatbib]{mn2e}
\usepackage{amssymb}
\usepackage{amsmath}
\usepackage[dvips]{graphicx}
\usepackage{epsfig}
\usepackage{multicol}

\date{Accepted ... Received ...; in original form ...}
\title{Jet simulations and Gamma-ray burst afterglow jet breaks}
\author[Van Eerten et al.]{H.J. van Eerten$^{1,2,5}$\thanks{E-mail: H.J.vanEerten@uva.nl}, Z. Meliani$^2$, R.A.M.J. Wijers$^1$, R. Keppens$^{2,3,4}$\\
$^{1}$Astronomical Institute 'Anton Pannekoek', PO box 94248, 1090 SJ Amsterdam, the Netherlands\\
$^{2}$Centre for Plasma Astrophysics, K.U. Leuven, Celestijnenlaan 200B, 3001 Leuven, Belgium\\
$^{3}$FOM-Institute for Plasma Physics Rijnhuizen, Nieuwegein, The Netherlands\\
$^{4}$Astronomical Institute, Utrecht University, The Netherlands\\
$^{5}$Center for Cosmology and Particle Physics, Physics Department, New York University, New York, NY 10003}

\begin{document}

\maketitle

\begin{abstract}
The conventional derivation of the gamma-ray burst afterglow jet break time uses only the blast wave fluid Lorentz factor and therefore leads to an achromatic break. We show that in general gamma-ray burst afterglow jet breaks are chromatic across the self-absorption break. Depending on circumstances, the radio jet break may be postponed significantly. Using high-accuracy adaptive mesh fluid simulations in one dimension, coupled to a detailed synchrotron radiation code, we demonstrate that this is true even for the standard fireball model and hard-edged jets. We confirm these effects with a simulation in two dimensions. The frequency dependence of the jet break is a result of the angle dependence of the emission, the changing optical depth in the self-absorbed regime and the shape of the synchrotron spectrum in general. In the optically thin case the conventional analysis systematically overestimates the jet break time, leading to inferred opening angles that are underestimated by a factor 1.32 and explosion energies that are underestimated by a factor 1.73, for explosions in a homogeneous environment.
\end{abstract}

\section{Introduction}

Gamma-ray burst (GRB) afterglows are produced by jetted outflows slowing down from ultrarelativistic velocities to non-relativistic velocities. Synchrotron emission results when the blast wave interacts with the circumburst medium and electrons get shock-accelerated and are accelerated in magnetic fields that are also usually presumed to be generated at the shock.

The jet nature of afterglows is not observed directly (there are no spatially resolved jet images). Theoretically, the energetics of the fireball model require that the outflow be concentrated towards the observer, since a spherical outflow would imply unphysically high explosion energy on the order of the Solar rest mass ($\backsim 10^{54}$ erg) instead of a more reasonable $\backsim 10^{52}$ erg. Observationally, afterglow light curves at various frequencies show a break followed by a steepening of the curve. This can be explained as a \emph{jet break}, beyond which relativistic beaming effects have decreased in strength to an extent where it becomes possible to distinguish between a jet and spherical outflow. A lack of emission from beyond the jet opening angle is now seen. A second effect that is theorized to occur at roughly the same time is a diminishing of synchrotron emission due to the onset of significant lateral expansion of the jet.

The conventional theoretical argument for the appearance of a jet break compares the \emph{half opening angle} $\theta_\mathrm{h}$ of the jet to the fluid Lorentz factor $\gamma$ directly behind the shock front. Once the latter has decreased to $\gamma \backsim 1 / \theta_\mathrm{h}$, beaming has sufficiently decreased and the absence of emission from angles beyond $\theta_\mathrm{h}$ becomes noticable. Since this ignores the basic shape of the synchrotron spectrum as well as electron cooling and the increase in optical depth due to synchrotron self-absorption, this suggests an \emph{achromatic} break in the light curve.

However, achromatic jet breaks are relatively rare in the literature. Observations in the \emph{Swift} era often do not show a clear jet break in the X-ray when the optical light curve does show a break (e.g. GRB990510, GRB030329, GRB060206). Also, the jet opening angle inferred from radio observations is often much larger than that from (earlier) X-ray or optical observations, which is again at odds with the expected achromaticity of the break.

Different explanations for the lack of truly achromatic breaks have been put forth in the literature. Staying close to the data, first of all \cite{Curran2008} show that current observations do not actually rule out an achromatic break between X-ray and optical and that both single and broken power law fits are often consistent with the X-ray data. On the theoretical side, the afterglow jet has been expanded into a structured jet (e.g. \citealt{Rossi2002}) or a superposition of multiple hard-edged jets with different opening angle. By now superpositions of up to three jets have been used to fit the data of GRB030329, with the opening angle explaining the radio light curve significantly larger than the others \citep{vanderHorst2005}. An overview of afterglow jet structure and dynamics can be found in \cite{Granot2007}.

Oddly enough, the analytical argument leading to an achromatic break has never been challenged. In this paper we present results from high resolution numerical jet simulations, coupled to an advanced synchrotron radiation code, showing that the standard analytical argument systematically overestimates the jet break time.

We explain the numerical set-up of the simulations in this paper in section \ref{set_up_section}. In section \ref{optically_thin_section} we study jet breaks while ignoring both self-absorption and cooling and find that the basic calculation linking jet break time and opening angle can be off by days. We derive a relation between jet break time and opening angle in the optically thin case and compare limb-brightening of the afterglow image at different spectral regimes.

In section \ref{jetbreaks_full_section} we calculate the full synchrotron spectrum and compare jet breaks at different spectral regimes. We find that the break at radio frequencies is postponed compared to the jet breaks observed at frequencies above the self-absorption break frequency. Depending on opening angle and observer frequency, this time difference can be on the order of several days (over a factor 2 in jet break time), even though we did not add any novel radiation physics or make any nonstandard assumptions. The new aspects of our calculation are merely the accuracy of the radiation code and the numerical resolution of the fluid simulation. Self-absorption is fully treated using linear radiative transfer equations and local electron cooling times are numerically calculated through an advection equation. The difference in jet break characteristics can be understood from the fact that different regions of the jet provide the dominant contribution for different observer frequencies. We show images of the emission coefficient throughout the blast wave to visualise the underlying physics.

Up to section \ref{jetbreaks_full_section} we assume that collimated outflow can be represented by a conic section from a 1D spherically symmetric simulation. We test this assumption, which implies that lateral spreading has little effect on the observed jet break, by performing a 2D simulation in section \ref{2D_simulation_section}. Higher dimensional simulations are not the focus of this paper and we will only briefly discuss the consequences of lateral spreading. We end with a summary and discussion of our results in section \ref{jetbreaks_summary_section}.

\section{Simulation and physics settings}
\label{set_up_section}

We have performed one-dimensional blast wave simulations using the relativistic hydrodynamics module of the \textsc{amrvac} adaptive-mesh refinement magnetohydrodynamics code (\citealt{Keppens2003, Meliani2007}). We have used an advanced equation of state (EOS) that implements an effective adiabatic index that gradually changes from $4/3$ in the relativistic regime to $5/3$ in the nonrelativistic regime. The upper cut-off Lorentz factor $\gamma_M$ of the shock-accelerated electron power law distribution is set to a numerically high value upstream and traced locally using an advection equation. This cut-off determines the position of the cooling break $\nu_c$ in the spectrum. The application of the advection equation and the EOS are introduced and explained in \cite{vanEerten2009c} (VE09). Some modifications to this method are explained in appendix \ref{numerical_approach_section}. We assume that little lateral spreading of the jet has taken place, and that a collimated outflow can be adequately represented by a conic section of a spherically symmetric simulation. The settings for the 2D simulation used to test this assumption are discussed separately in section \ref{2D_simulation_section}.

We have used the following physics settings: explosion energy $E = 2.6 \cdot 10^{51}$ erg, (homogeneous) circumburst number density $n_0 = 0.78$ cm$^{-3}$, accelerated electron power law slope $p = 2.1$, fraction of thermal energy density in the accelerated electrons $\epsilon_E$ and in the magnetic field $\epsilon_B$ both equal to 0.27 and the fraction $\xi_N$ of electrons accelerated at the shock front equal to 1.0 (unless explicitly stated otherwise). Unlike in the simulations performed in VE09, we have kept the fractions $\xi_N$, $\epsilon_E$ and $\epsilon_B$ fixed throughout the simulation, in order to stay as close as possible to the conventional fireball model. We have set the observer luminosity distance $r_{obs} = 2.47 \cdot 10^{27}$ cm, but kept the redshift at zero (instead of the matching value 0.1685). The observer is assumed to be positioned on the axis of the jet. These physics settings qualitatively describe GRB030329 and are identical to those used in VE09. They do not provide a quantitative match to the data for GRB030329 however, since they have been derived using an analytical model by \cite{vanderHorst2008mar} and not by directly matching simulation results to observational data.

We have run a number of simulations, using a grid with 10 base blocks (of 12 cells each) and up to 19 refinement levels (with the resolution doubling at each next refinement level) that has boundaries at $1.12 \cdot 10^{14}$ cm and $1.12 \cdot 10^{18}$ cm. A blast wave reaching the outer boundary provides coverage up to an observation time of $\backsim 50$ days. We start the simulation with a blast wave with shock Lorentz factor 25, ensuring complete coverage long before one day in observer time. We have checked that this resolution is sufficient and discuss this in appendix \ref{numerical_approach_section}.

The synchrotron radiation has been calculated with the method introduced in \cite{vanEerten2009} and VE09, using a linear radiative transfer method where the number of calculated rays is changed dynamically via a process analogous to adaptive mesh refinement for the RHD simulation. We have allowed for 19 refinement levels in the radiation calculation, just like in the RHD simulation. The radiative transfer calculation uses the output from the dynamics simulation that has been stored at fixed time intervals, and we have used up to 10,000 such snapshots of the fluid state.

\section{Optically thin blast waves and break times}
\label{optically_thin_section}
\begin{figure}
\centering
\includegraphics[width=1.0\columnwidth]{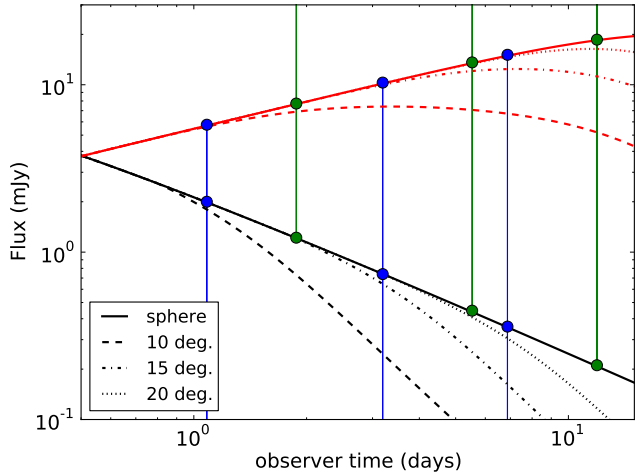}
\caption{Light curves without self-absorption and electron cooling, leaving only critical frequency $\nu_m$. The light curves are calculated at observer frequencies $1.4 \cdot 10^9$ Hz (upper light curves) and $5 \cdot 10^{17}$ Hz (lower light curves), to ensure they are on different sides of $\nu_m$. Results for $5 \cdot 10^{17}$ Hz have been multiplied by a scaling factor 100. Aside from results from spherical flow, results from hard-edged jets with half opening angles of 10, 15 and 20 degrees are plotted. The new predicted jet break times are indicated by dots connected to the lower edge of the plot, old predictions connect to the upper edge of the plot.}
\label{break_curves_nosanoc_figure}
\end{figure}
Already in a simplified set-up where we ignore self-absorption and electron cooling we find a noticable difference between the analytically expected jet break time and the simulation results. In figure \ref{break_curves_nosanoc_figure} we have plotted light curves at two different frequencies, $1.4 \cdot 10^9$ Hz and $5 \cdot 10^{17}$ Hz, chosen such that they lie safely below and above the synchrotron peak frequency $\nu_m$. Also indicated in the plot are both the conventional estimate for the jet break time and an improved estimate (both explained below).  

The jet break time is conventionally linked to the jet half opening angle $\theta_h$ using the Blandford-McKee (BM) self-similar solution for the blast wave dynamics in the ultrarelativistic regime \citep{Blandford1976}. The argument is as follows. According to BM the blast wave shock Lorentz factor $\Gamma$ and the emission time $t_e$ are related via
\begin{equation}
\Gamma^2 = \frac{E (17-4k)}{8 \pi m_p n_0 R_0^k c^{5-k}} t_e^{k-3} \equiv A_1 t_e^{k-3},
\end{equation}
where $m_p$ the proton mass, $n_0$ the proton number density at distance $R_0$, $c$ the speed of light and $k$ the slope of the circumburst medium number density $n$ ( with $n(r) \equiv n_0 ( r / R_0 )^{-k}$). The position of the shock front $R$ is given by
\begin{equation}
 R = c t_e ( 1 - \frac{1}{2(4-k) \Gamma^2} ),
\end{equation}
and the shock Lorentz factor and the fluid Lorentz factor at the shock front $\gamma_\mathrm{f}$ are related via $\gamma_\mathrm{f}^2 = \Gamma^2 / 2$. From the relation between observer time and emission time for the shock front,
\begin{equation}
 t_\mathrm{obs} = t_e - \cos \theta_\mathrm{h} R / c,
\end{equation}
and the half opening angle of the relativistic beaming cone $\theta = 1 / \gamma$, it now follows that
\begin{equation}
 t_\mathrm{obs,break} = \theta_\mathrm{h}^{2+2/(3-k)} \cdot \left( \frac{A_1}{2} \right)^{1/(3-k)} \cdot ( \frac{1}{2} + \frac{1}{4(4-k)} ).
\label{break_time_front_equation}
\end{equation}

When we take the radial profile of the emitting region into account, the jet break can be estimated somewhat more accurately. Although even in the optically thin regime, the emission from the shock front dominates the total emission, we start noticing a lack of emission from the back of the blast wave, since there the fluid Lorentz factor is the smallest and the corresponding relativistic cone the widest. The width of the blast wave is approximately $R / \Gamma^2$. The fluid Lorentz factor $\gamma_\mathrm{b}$ at this position is approximately given by
\begin{equation}
\gamma_\mathrm{b}^2 \equiv \left( 1 + 2(4-k) \right)^{-1} \Gamma^2 / 2.
\end{equation}
If we now use
\begin{equation}
 t_\mathrm{obs} = t_e - \cos \theta_\mathrm{h} R ( 1 - 1 / \Gamma^2 ) / c,
\end{equation}
we find
\begin{equation}
\begin{array}{rl}
 t_\mathrm{obs,break} & = \theta_\mathrm{h}^{2+2/(3-k)} \cdot \left( \frac{A_1}{2(1+2(4-k))} \right)^{1/(3-k)} \times \\
  & \left( \frac{1}{2} + \frac{1}{2(1+2(4-k))} + \frac{1}{2(4-k)2(1+2(4-k))} \right).
\end{array}
\label{break_time_back_equation}
\end{equation}

The comparison between the analytically expected break times using the shock front and the actual break times from the simulations in fig. \ref{break_curves_nosanoc_figure} shows that the expected break times systematically overestimate the real break times, whereas jet break times estimated using the back of the blast wave lie consistently closer to the real break times. Jet break predictions using the blast wave back for 10, 15, 20 deg. half opening angle are 1.1, 3.2, 6.9 days respectively. Using the blast wave front we find 1.9, 5.5, 12 days instead. The precise value of the real break times depend on how this term is defined. The improved estimates are closer when the break time is defined as the meeting point of the pre- and post-break asymptotes (which is in practice easier to determine for observer frequencies above $\nu_m$ because then the transition between the regimes and the corresponding change in temporal slope have already occured). The improved estimates are also closer when approximating the break time to be where the spherical and collimated outflow curves start to differ noticably, although some divergence between the two can already be seen before the improved break time estimations. This is not unexpected since even the improved jet break time estimate is inexact, depending on approximating gradually changing features like the edge of the beaming cone and the back of the blast wave by sudden transitions. When a broken power law is fit to observational data to determine the jet break time, this time is usually defined as the meeting point of the asymptotes.

For a homogeneous circumstellar environment, the difference between basic and improved observer time estimates is a fixed factor of roughly one half (0.48), with the conventional analysis overestimating the jet break time. The corresponding correction factor to correct jet opening angle estimates that have been obtained using the blast wave front (and therefore underestimated), is therefore 1.32. The underestimated explosion energies require a correction factor 1.73. 

\begin{figure}
\centering
\includegraphics[width=1.0\columnwidth]{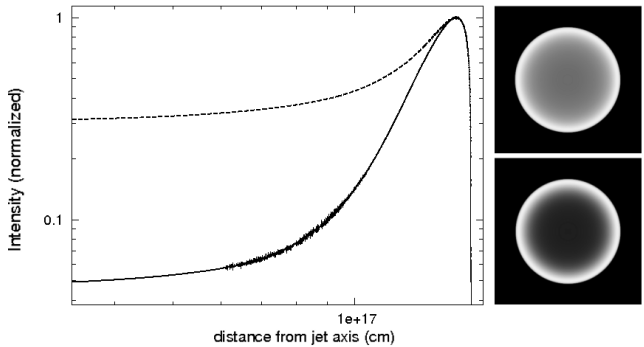}
\caption{Spatially resolved images both above $\nu_m$ (solid line) and below $\nu_m$ (dashed line), for spherical explosions around ten days observer time. For the lower frequency $1.4\cdot10^9$ Hz has been used, and for the higher frequency $5 \cdot 10^{17}$ Hz has been used. Self-absorption and electron cooling are excluded. Top right image is below $\nu_m$, bottom right image is above $\nu_m$. Although both images are limb-brightened, the profile is sharper for the high frequency, which explains why the jet break is sharper at the high frequency.}
\label{spatially_resolved_figure}
\end{figure}
Jet breaks at frequencies below $\nu_m$ are less sharp and therefore may appear to occur later. This effect can be attributed to the different level of limb-brightening at both sides of the $\nu_m$. Spatially resolved images at 10 days for both frequencies are shown in figure \ref{spatially_resolved_figure}. When the intensity peaks at the same radius in the image, but the decline of the intensity is less steep moving to lower radii, the jet break will be more gradual as well.

\section{jet breaks at different frequency ranges}
\label{jetbreaks_full_section}
\begin{figure}
\centering
\includegraphics[width=1.0\columnwidth]{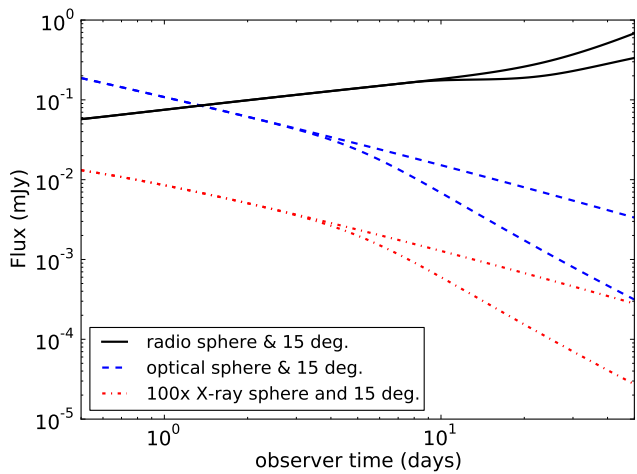}
\caption{Light curves between 0.5 and 28 days in observer time for a radio ($1.4 \cdot 10^{9}$ Hz), optical ($5 \cdot 10^{14}$ Hz) and X-ray ($5 \cdot 10^{17}$ Hz) frequency, in each case both for a spherical explosion and a hard-edged jet with half opening angle of 15 degrees. Both the optical and X-ray curve lie above the cooling break $\nu_c$. The radio jet break is postponed with respect to the others, at least by several days. The radio light curves change to a steeper rise around 25 days because $\nu_m$ is crossed before $\nu_a$.}
\label{lightcurves_main_figure}
\end{figure}
\begin{figure}
\centering
\includegraphics[width=1.0\columnwidth]{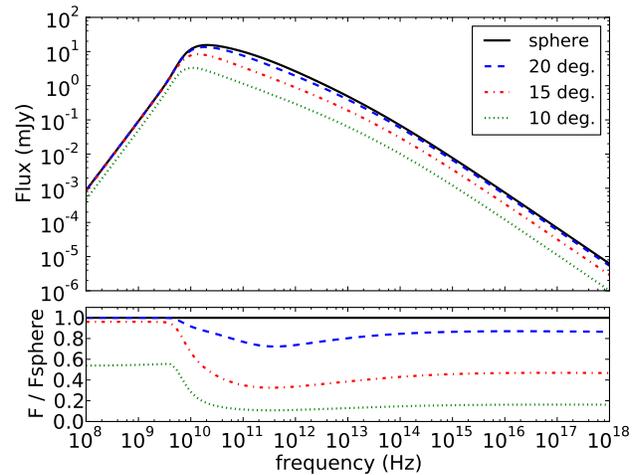}
\caption{Spectra at ten days observer time for a spherical explosion and various hard-edged jet half-opening angles. The lower plot shows the flux for the same spectra, now as fraction of the spherical case. Only the 10 degrees half opening angle jet differs noticably from the spherical case at radio frequencies.}
\label{jetbreak_spectra_figure}
\end{figure}

In figure \ref{lightcurves_main_figure} we show light curves calculated using the full synchrotron spectrum. The main result here is that the jet break below the self-absorption break $\nu_a$ is postponed by several days with respect to the others. The chromaticity of the jet break is made explicit in figure \ref{jetbreak_spectra_figure}, showing the spectrum for a spherical outflow and collimated outflows with varying opening angles at 10 days in observer time. Below both $\nu_m$ and $\nu_{a}$ the dinstinction between the flux levels for the different opening angles becomes a lot smaller. The flux for 20 degrees is even indistinguishable from the spherical case, implying that there is no jet break yet in the radio, while it has already occurred at higher frequencies. For a half opening angle of 15 degrees, the break has only barely set in at ten days in the radio.

\begin{figure}
\centering
\includegraphics[width=1.0\columnwidth]{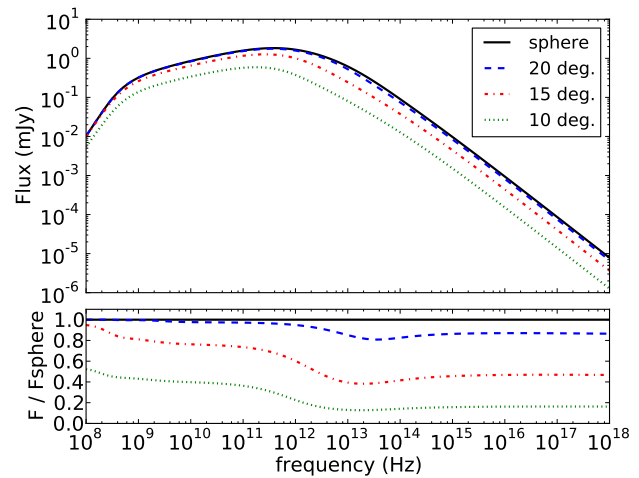}
\caption{Spectra at ten days observer time for a spherical explosion and various hard-edged jet half-opening angles, with $\xi_N = 0.1$. The lower plot shows the flux for the same spectra, now as fraction of the spherical case.}
\label{jetbreak_spectra_xi01_figure}
\end{figure}
\begin{figure}
\centering
\includegraphics[width=1.0\columnwidth]{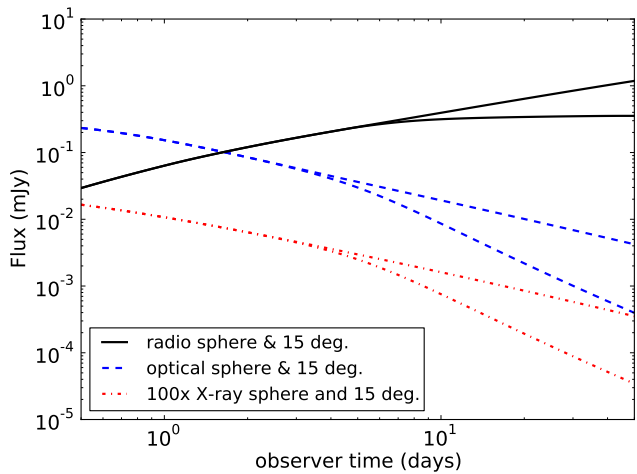}
\caption{Light curves between 0.5 and 28 days in observer time for a radio ($1.4 \cdot 10^{9}$ Hz), optical ($5 \cdot 10^{14}$ Hz) and X-ray ($5 \cdot 10^{17}$ Hz) frequency, with $\xi_N = 0.1$.}
\label{lightcurves_xi01_main_figure}
\end{figure}
For the settings that we have used so far, $\nu_m$ and $\nu_a$ lie very close together at 10 days observer time. In order to make the difference between the two clear and show the spectral regime in between, we have calculated spectra at ten days using $\xi_N = 0.1$ as well. That we have chosen to alter this parameter instead of any of the others carries no special significance, but merely serves to move $\nu_m$ and $\nu_a$ apart. Figure \ref{jetbreak_spectra_xi01_figure} shows the resulting spectra. As indicated by the lower plot, the radio light curve at $1.4$ GHz now lies above $\nu_a$ and no longer leads to a postponed break time for a half opening angle of 15 degrees. Fig. \ref{lightcurves_xi01_main_figure} shows the light curves for $\xi_N = 0.1$. The radio curve lies above the self-absorption break and the radio jet break has now moved close to the other breaks. There is still no readily discernable difference between the optical and X-ray jet break times.

\begin{figure}
\centering
\includegraphics[width=0.48\columnwidth]{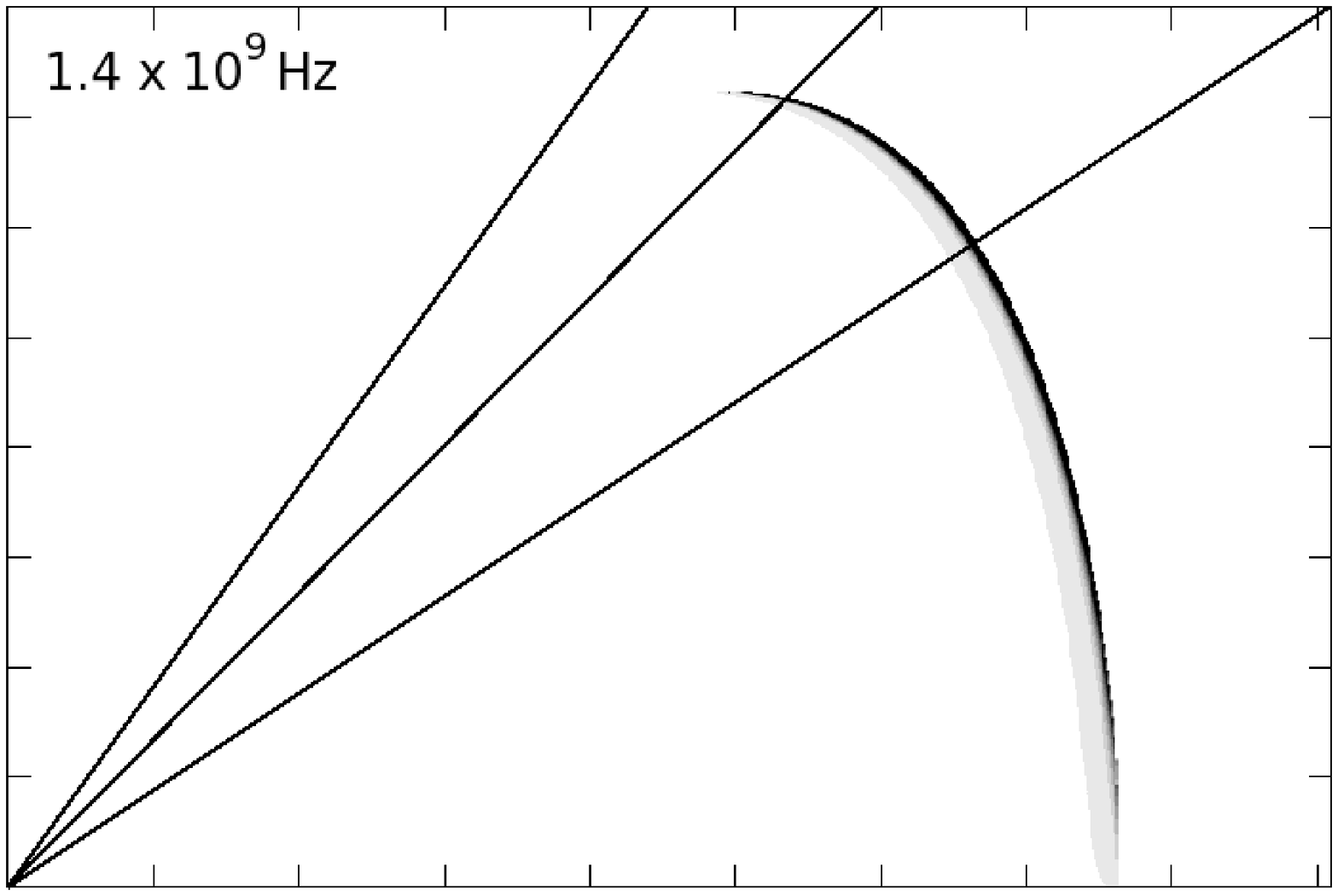}
\includegraphics[width=0.48\columnwidth]{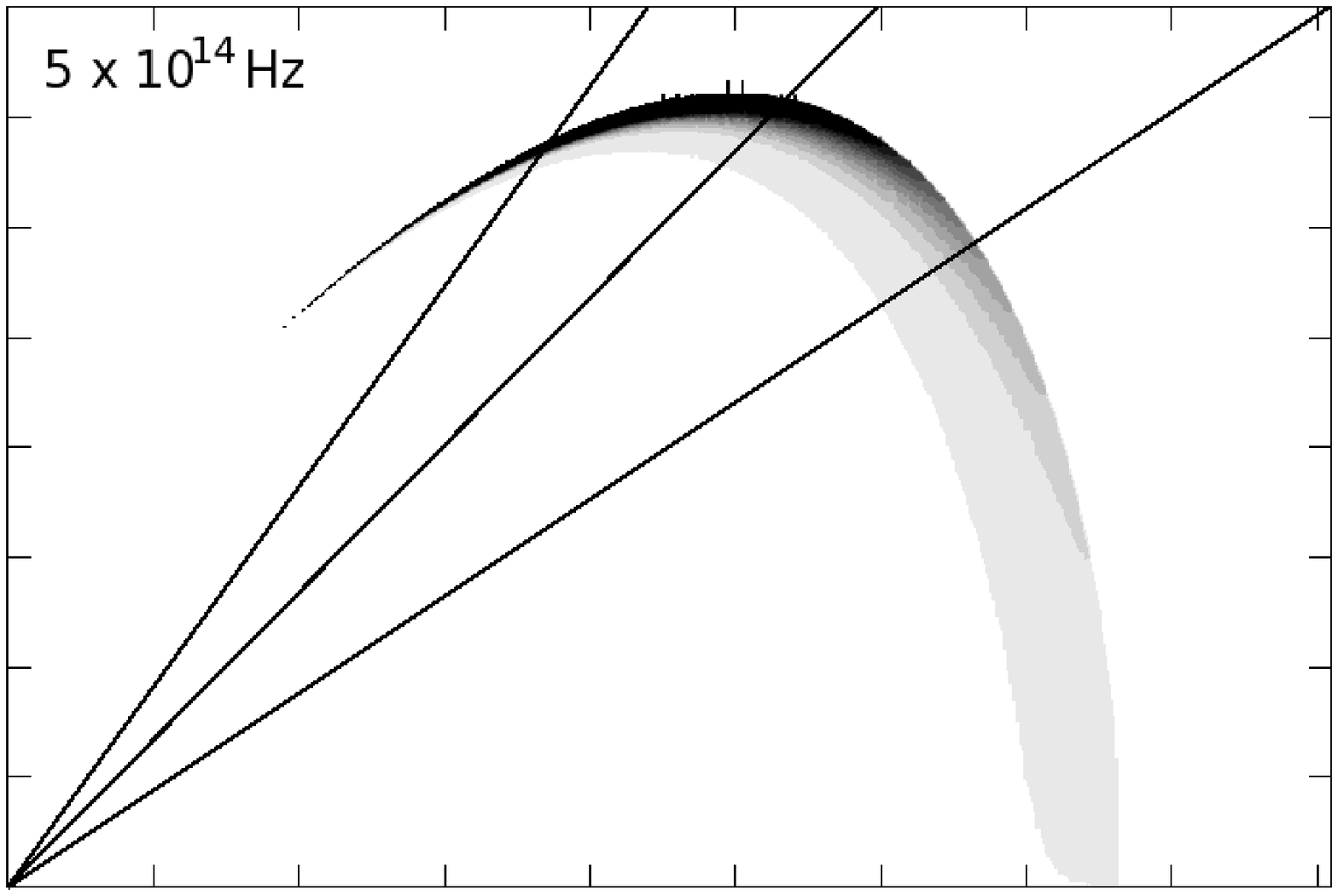}
\includegraphics[width=0.48\columnwidth]{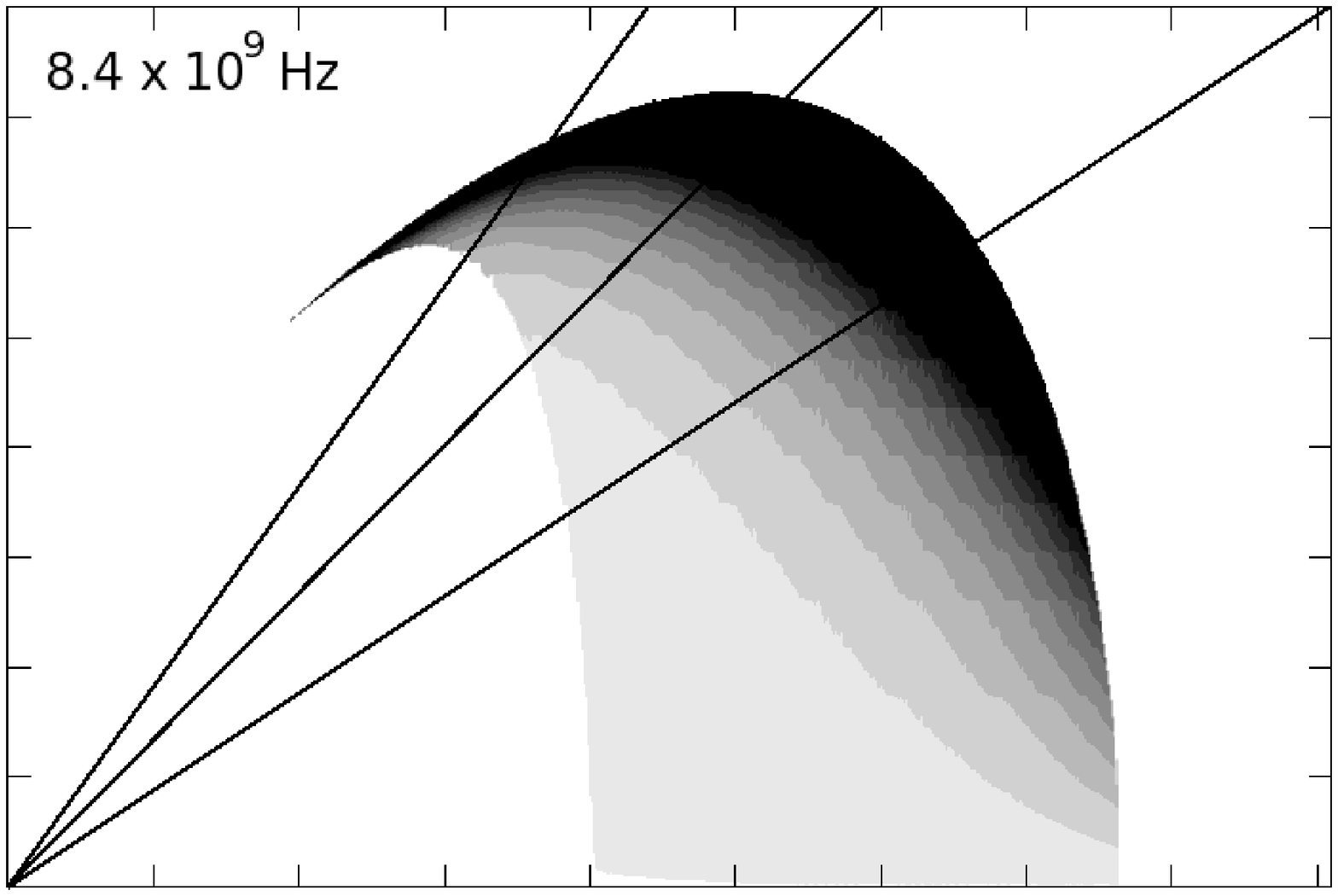}
\includegraphics[width=0.48\columnwidth]{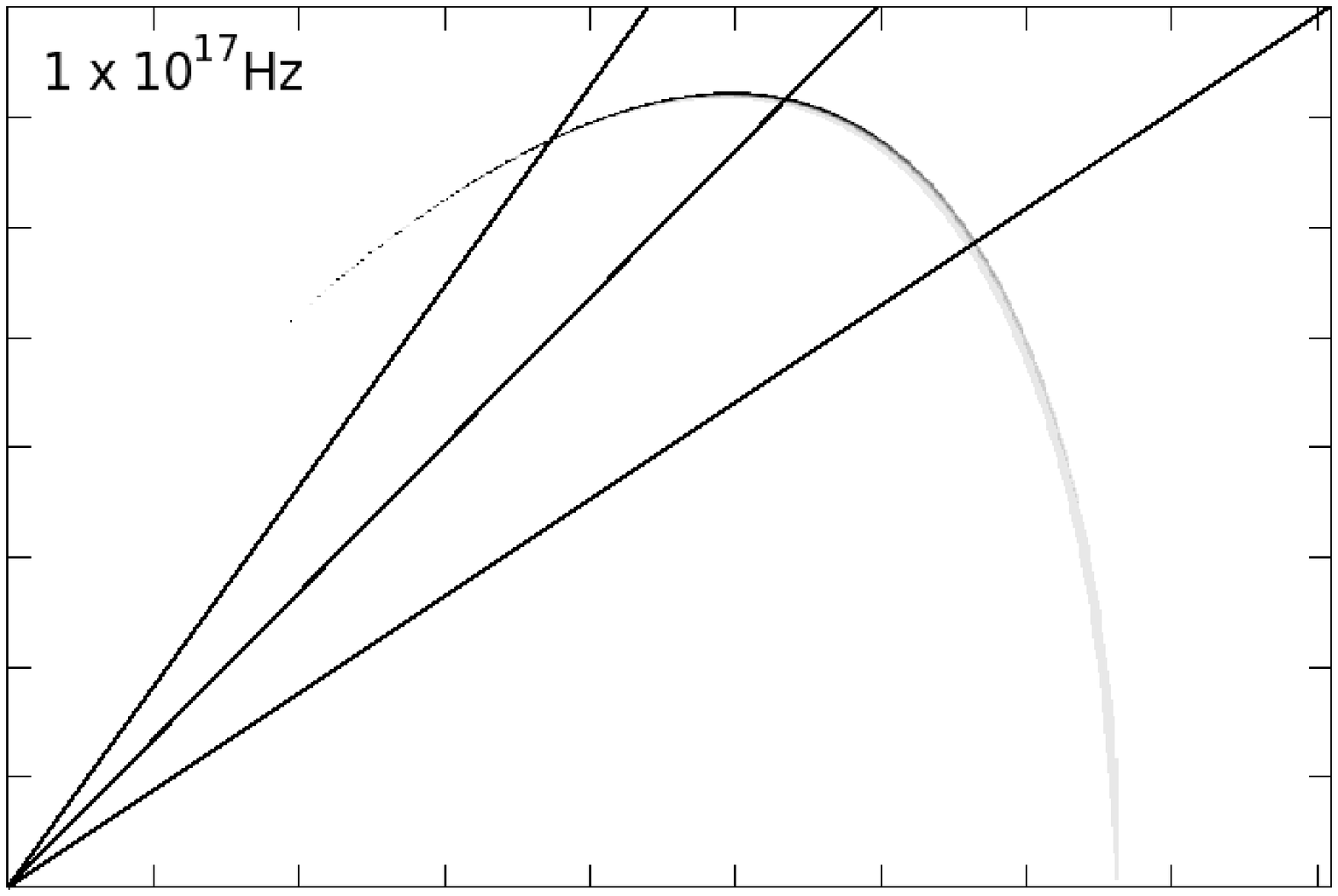}
\includegraphics[width=0.48\columnwidth]{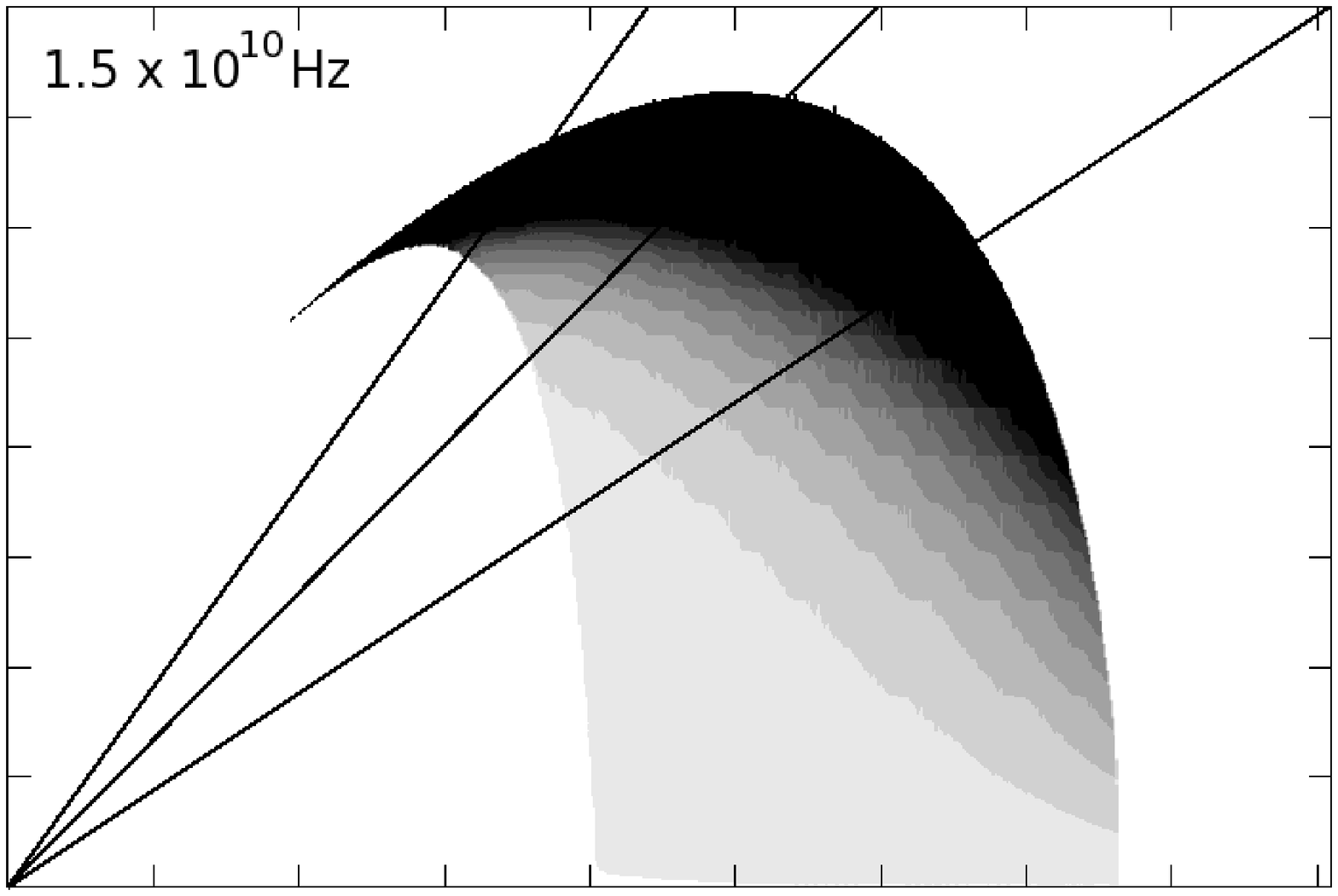}
\includegraphics[width=0.48\columnwidth]{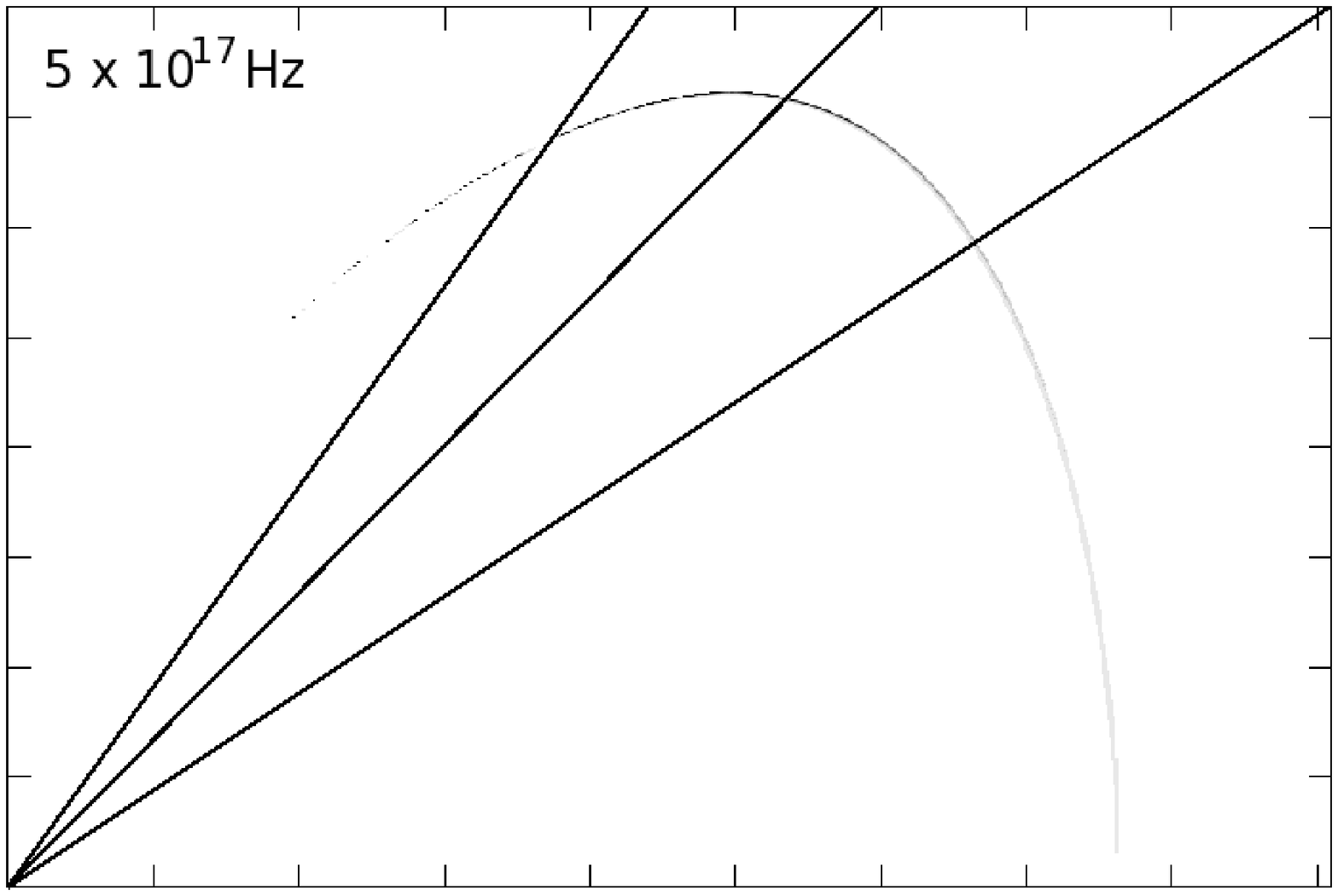}\caption{ring integrated, absorption corrected emission coefficients, for various frequencies and at 10 days in observer time. On the horizontal axis the position in cm in the $z$ direction (i.e. along the jet axis, the observer is located right of the plot) where the emission was generated, on the vertical axis the distance to the jet axis $h$ in cm. In the left column we have three radio frequencies: $1.4 \cdot 10^9$ Hz, $8.4 \cdot 10^9$ Hz, $1.5 \cdot 10^{10}$ Hz, top to bottom. In the right column we have optical and X-ray: $5 \cdot 10^{14}$ Hz, $1 \cdot 10^{17}$ Hz, $5 \cdot 10^{17}$ Hz, top to bottom. The horizontal scale is from 0 to $9 \cdot 10^{17}$ cm, and the vertical scale is from 0 to $1.6 \cdot 10^{17}$ cm, for each plot. The normalized grey-scale coding for each plot indicates the strength of emission. All emission coefficients are multiplied by $2 \pi h$, such that the plots show the proper relative contributions from the different angles. The emission coefficients are also corrected for optical depth. The diagonal lines denote jet half opening angles of 20, 15 and 10 degrees from left to right. For a given opening angle in the hard-edged jet case, everything above the diagonal is excluded.}
\label{emission_coefficients_figure}
\end{figure}

The physical mechanism behind the difference in jet breaks can be understood as follows. First, below the synchrotron break frequency $\nu_m$, the limb brightening becomes less strong and the main contributing region of the jet to the observed flux moves closer to the jet axis. Second, below the self-absorption break optical depth starts to play a role and the main contributing region moves even more towards the front of the jet. Both these effects move the contributing region of the jet away from the jet edge and closer to the center front of the jet. This effect is illustrated in figure \ref{emission_coefficients_figure}.

In these figures we have plotted ring-integrated, absorption corrected local emission coefficients for various frequencies at ten days observer time. The highlighted areas are the areas that, for the given frequency, contribute the most to the observed signal. All emission coefficients $j_\nu$ are multiplied by $2 \pi h$ (where $h$ is the distance to the jet axis), such that the plots show the proper relative contributions from the different angles. The emission coefficients are also corrected for optical depth ($\tau$), so the quantity that is plotted is $j_\nu 2 \pi h \exp[ - \tau ]$. The diagonal lines denote jet half opening angles of 20, 15 and 10 degrees from left to right (the shape of lines along a fixed angle is not affected by the transition from emission to observer frame, this only introduces a transformation \emph{along} the radial lines). For a given opening angle in the hard-edged jet case, everything above the diagonal is excluded. The plots immediately show why the jet break is postponed in the radio and differs in shape for different frequencies. 

Aside from showing the origin of the delay in the jet break for radio frequencies, the images also show us that for higher frequencies, we look at earlier emission times in general. From this it also follows that not just the jet break, but any variability resulting from changes in the fluid conditions, will likely manifest themselves in a chromatic fashion. The blast wave size is smaller at earlier times, and at early times, fluid perturbations will be less smeared out. Thus it follows that variability will be most clearly observed in the X-ray light curve.

As the bottom plots in figures \ref{jetbreak_spectra_figure} and \ref{jetbreak_spectra_xi01_figure} show, the fractional difference between spherical and collimated outflow is not entirely independent of frequency even above the self-absorption break. Both for $\xi_N = 0.1$ and $\xi_N = 1.0$, the collimated outflow flux over the spherical outflow flux reaches a minimum in the spectral region between $\nu_m$ and the cooling break $\nu_c$. This leaves open the possibility that for certain physics parameters and opening angles the jet break as inferred from observational data may differ between optical and X-ray, albeit with a difference that is far less pronounced than that across the self-absorption break that we have discussed above. A quantitative assessment of this effect can be made by including the observational biases and errors of measurements for the different frequencies as well, but lies outside the scope of this paper. Judging purely from the simulated light curves while assuming perfect coverage of the data, a strong distinction between optical and X-ray jet breaks is not obvious.

\section{self-absorption and jet break from a 2D simulation}
\label{2D_simulation_section}

\begin{figure}
\centering
\includegraphics[width=0.49\columnwidth]{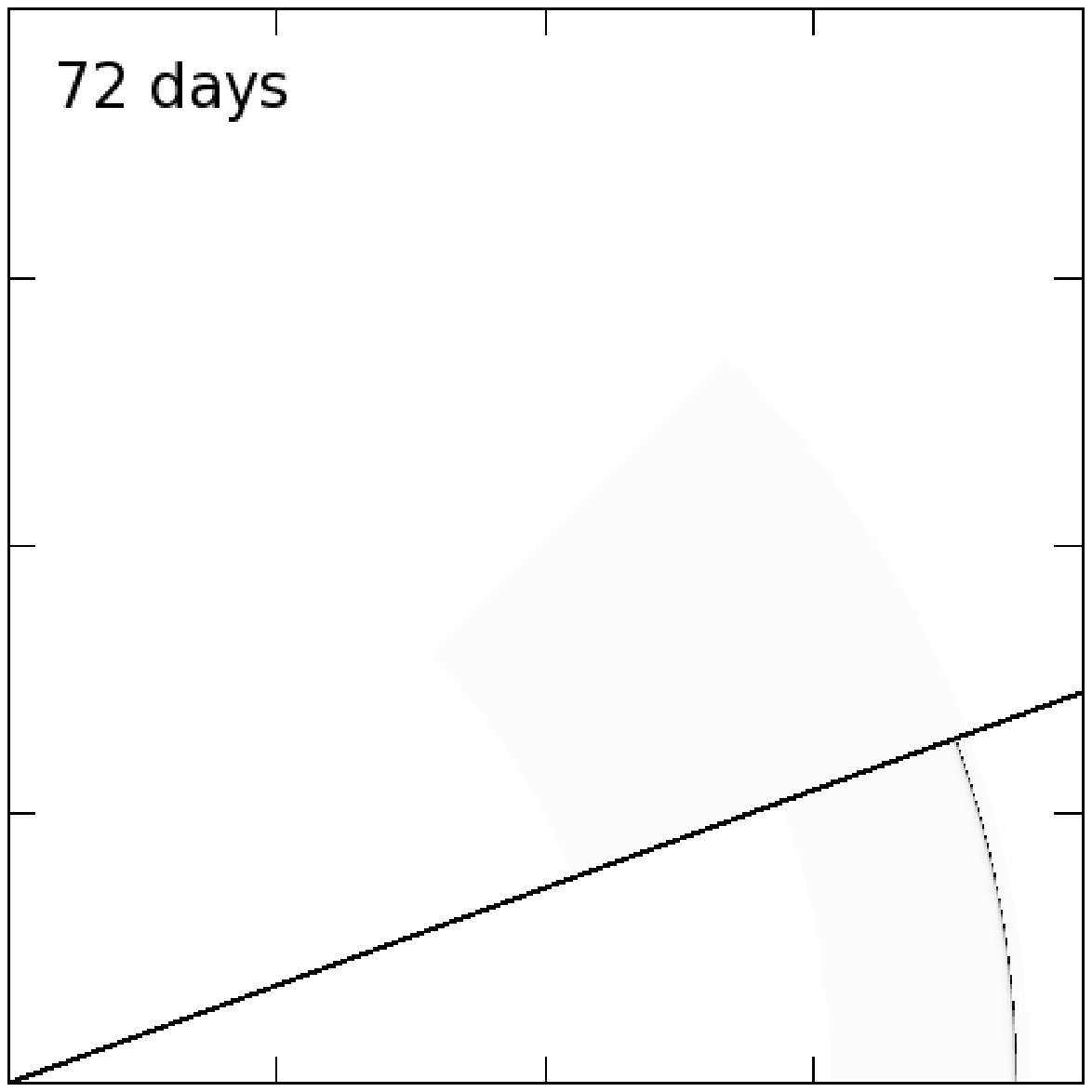}
\includegraphics[width=0.49\columnwidth]{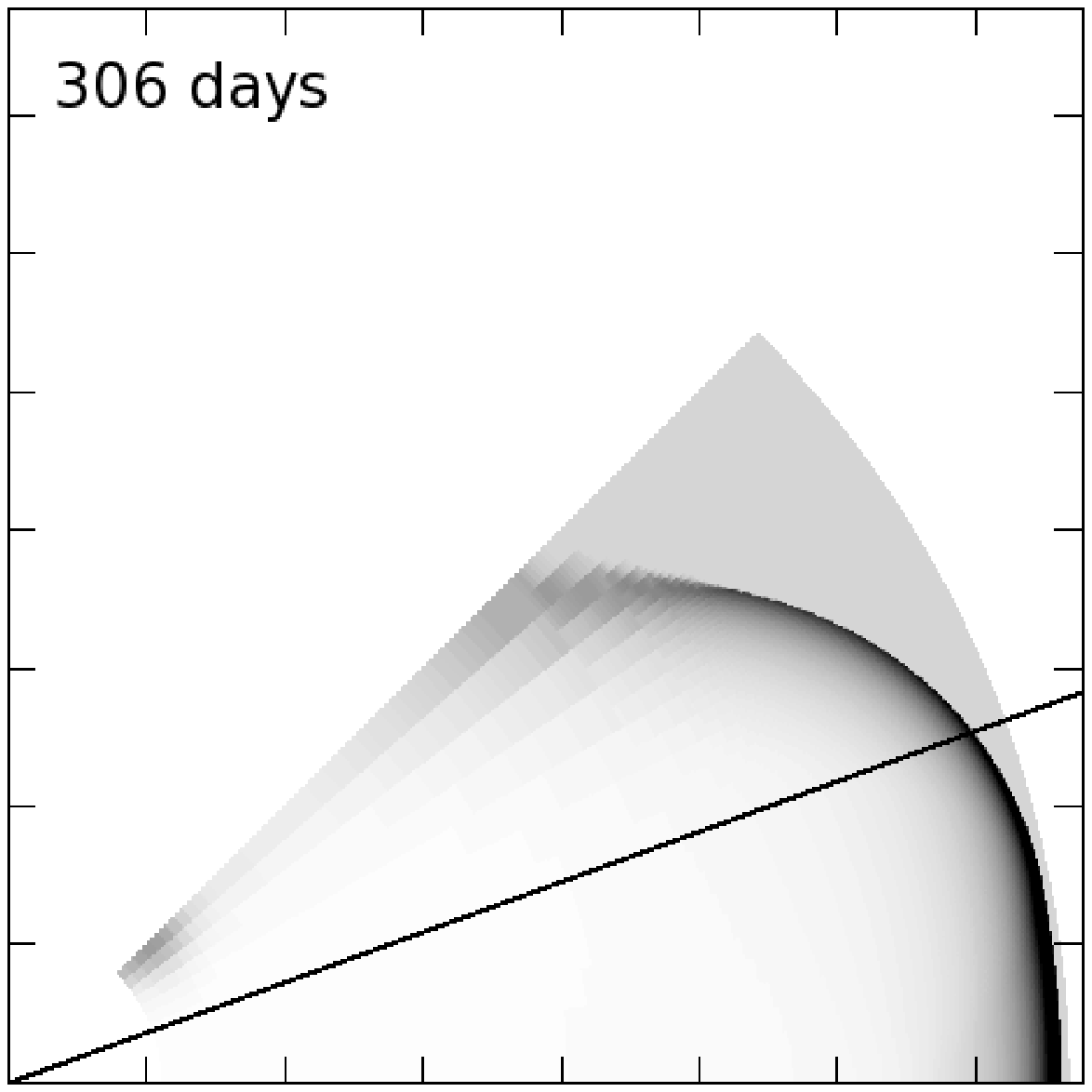}
\includegraphics[width=0.49\columnwidth]{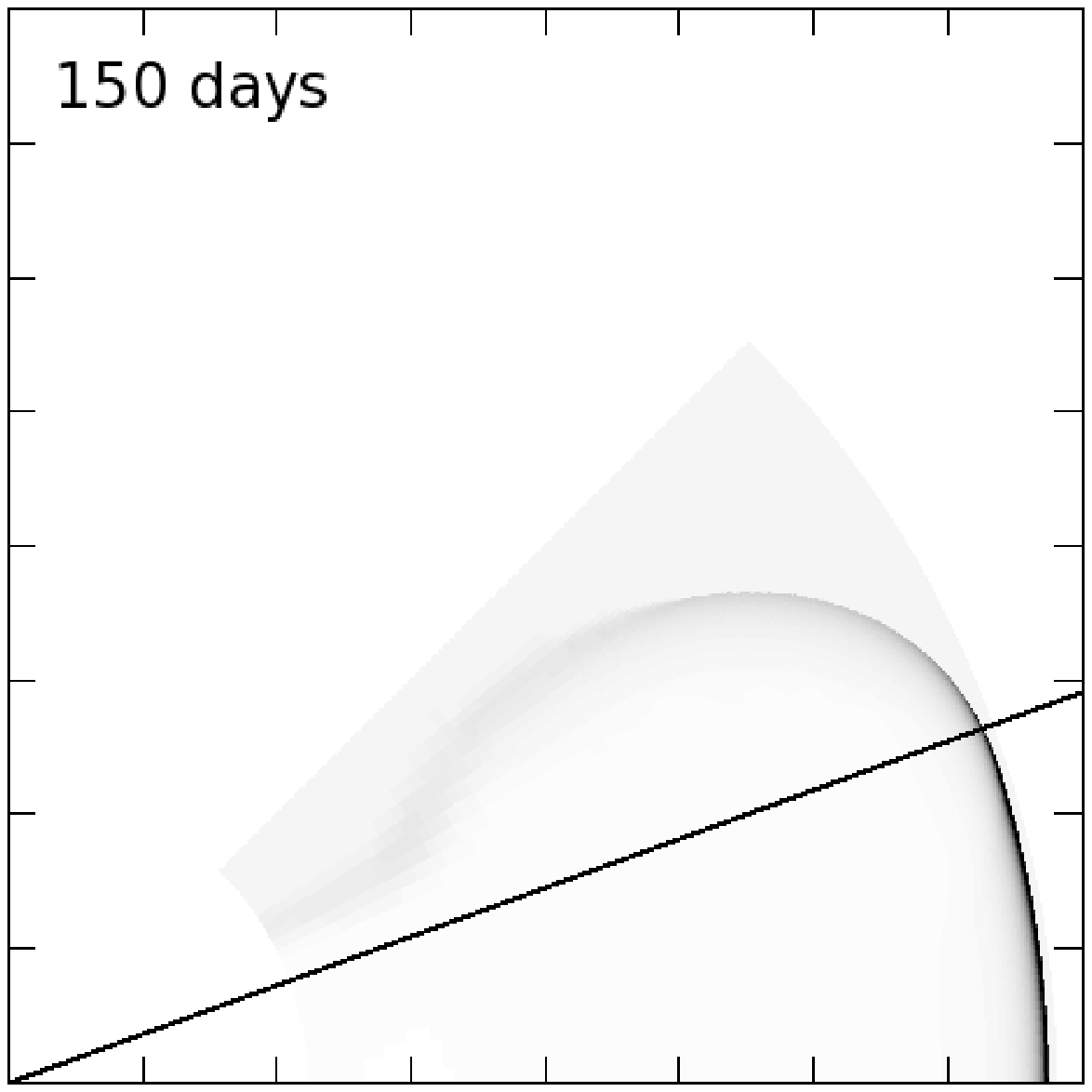}
\includegraphics[width=0.49\columnwidth]{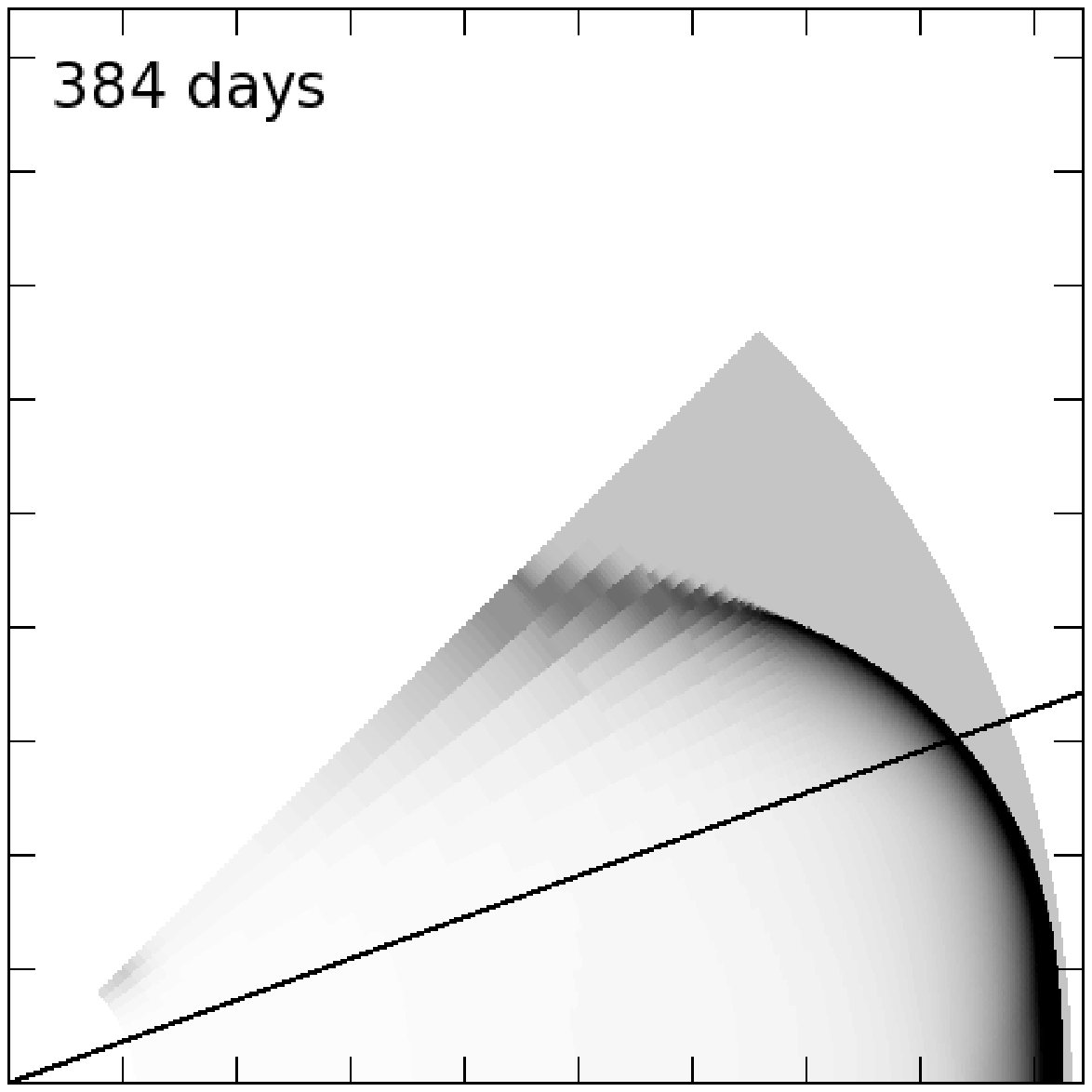}
\includegraphics[width=0.49\columnwidth]{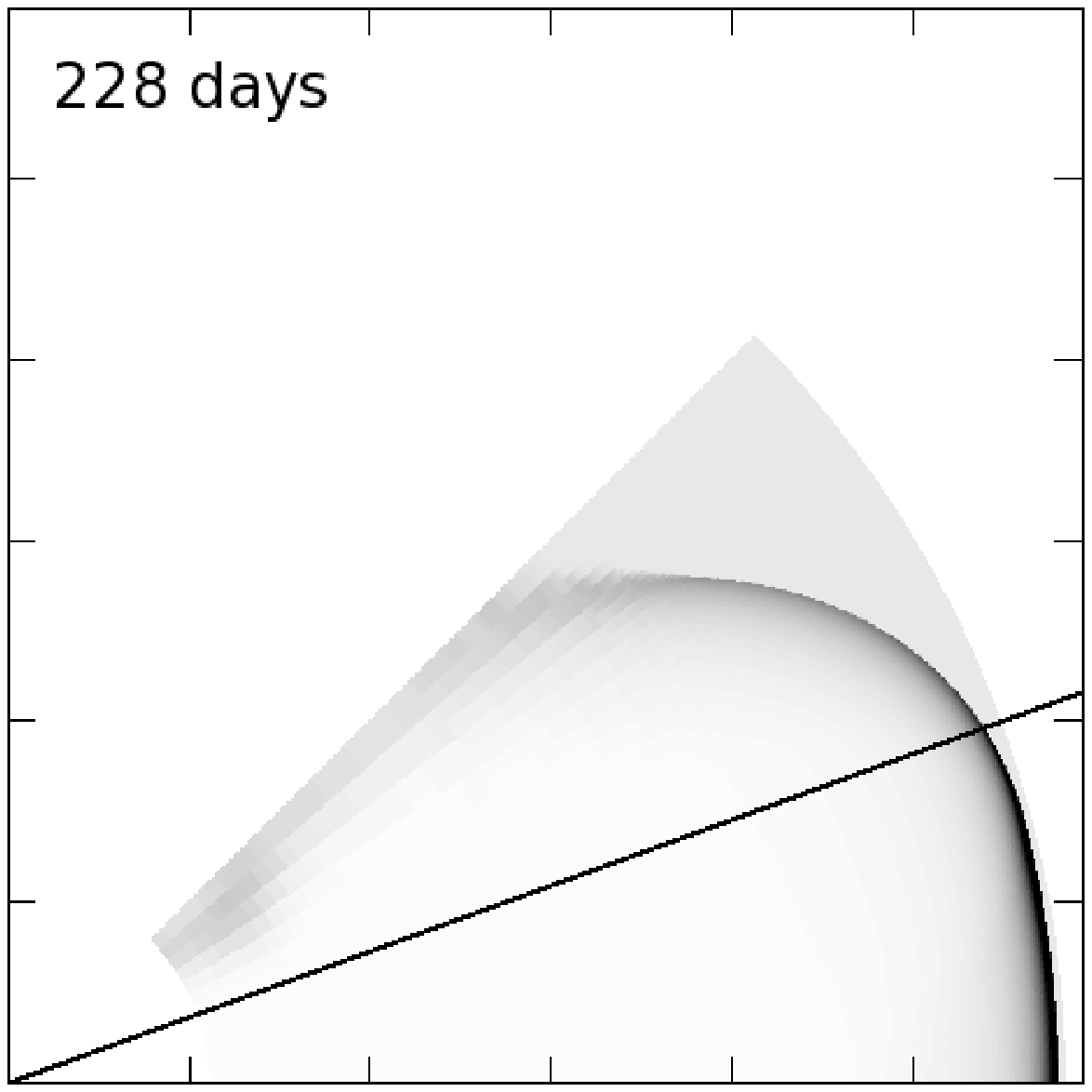}
\includegraphics[width=0.49\columnwidth]{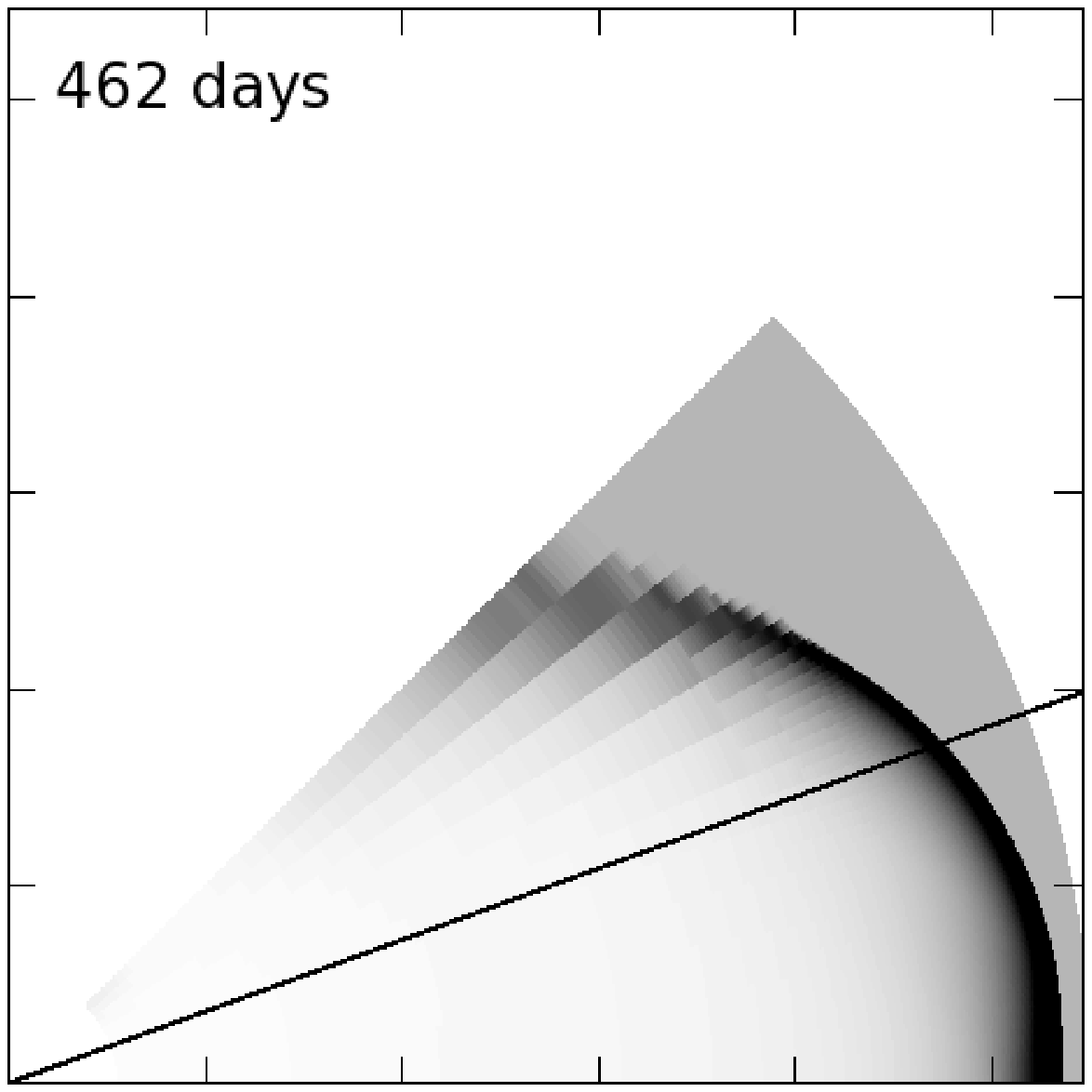}
\caption{Snapshots of lab frame density $D$ for the two-dimensional simulation. The grey scales are normalized with respect to the peak density in each snapshot. The axis are normalized as well. The starting half opening angle of 20 degrees is indicated by the diagonal line. For the left column the emission times (and blast wave radii) are 72 days ($1.88 \cdot 10^{17}$ cm), 150 days ($3.88 \cdot 10^{17}$ cm) and 228 days ($5.82 \cdot 10^{17}$ cm) from top to bottom. For the right column these are 306 days ($7.65 \cdot 10^{17}$ cm), 384 days ($9.31 \cdot 10^{17}$ cm) and 462 days ($1.08 \cdot 10^{18}$ cm). }
\label{dynamics_2D_figure}
\end{figure}

\begin{figure}
\centering
\includegraphics[width=1.0\columnwidth]{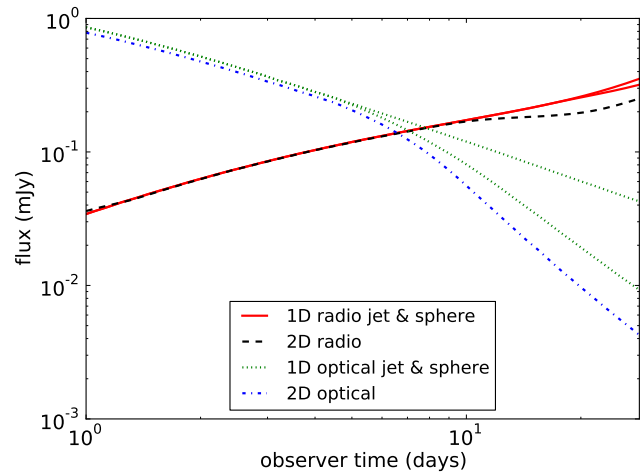}
\caption{Direct comparison between spherical explosion, a hard-edged jet of half opening angle 20 degrees and a two-dimensional simulation starting at half opening angle of 20 degrees. Light curves are shown for two frequencies: $1.4 \cdot 10^9$ Hz (radio) and $5 \cdot 10^{14}$ Hz (optical). Cooling effects are ignored. In the optical, the light curve from the 2D simulation results in a steeper break around the same time, while in the radio the jet break also occurs earlier than for the hard-edged jet model.}
\label{lightcurves_2D_figure}
\end{figure}

In order to confirm the chromaticity of the jet break in two dimensions we have run a simulation in 2D as well. We have used a similar set up as in the 1D case, starting with a hard-edged jet at Lorentz factor 15 with half opening angle of 20 degrees. In the angular direction we have used 1 base level block instead of ten (as in the radial direction). The maximum half opening angle covered for the jet is 45 degrees. For numerical reasons, the maximum refinement level is currently 11. This is sufficient to qualitatively capture the blast wave physics, but in order to draw more definitive conclusions a higher refinement level will be needed. Such simulations are currently being performed and will be presented in future work.

In fig. \ref{dynamics_2D_figure} we have plotted snapshots of the lab frame density structure of the jet for various emission times. They show lateral expansion of the jet. At this stage the lateral expansion is in excess of that predicted by \cite{Rhoads1999}, who predicted exponential expansion \emph{after} the jet has reached a certain radius ($1.28\cdot 10^{18}$ cm for our explosion parameters) and neglible expansion before. Once the lateral expansion supposedly sets in, this analytical estimate will quickly overestimate the jet opening angle. The fluid velocity at higher opening angles drops off quickly, as can be seen from the radius of the blast wave beyond the half opening angle of 20 degrees. This part of the fluid is not expected to contribute significantly to the observed emission. In order to speed up the calculation we have automatically derefined the grid both at very low densities and far downstream (using the Blandford-McKee solution to estimate the blast wave radius). The effect of this can be seen in the snapshot images at the back of the jet edge of the jet outside of the original jet opening angle. It has no effect on the light curve or the dynamics of the relevant part of the fluid. The precise dynamics of the two-dimensional afterglow blast wave will be discussed in future work. For now we note that our simulations show results consistent with \cite{Zhang2009}.

Fig. \ref{lightcurves_2D_figure} confirms the chromaticity of afterglow jet breaks in two dimensions. Because our approach to electron cooling requires a higher resolution than provided by 11 refinement levels, we have restricted ourselves to a comparison between the radio and optical light curves and the role of self-absorption. The first obvious result is that the 2D simulations indeed confirm the qualitative conclusion from 1D simulations and hard-edged jets that the jet break is chromatic. Also for 2D simulations, the radio jet break is postponed with respect to the optical jet break. The second result is that the effect of lateral spreading on the radio jet break partially (but not completely) counteracts the delay in jet break.

\begin{figure}
\centering
\includegraphics[width=1.0\columnwidth]{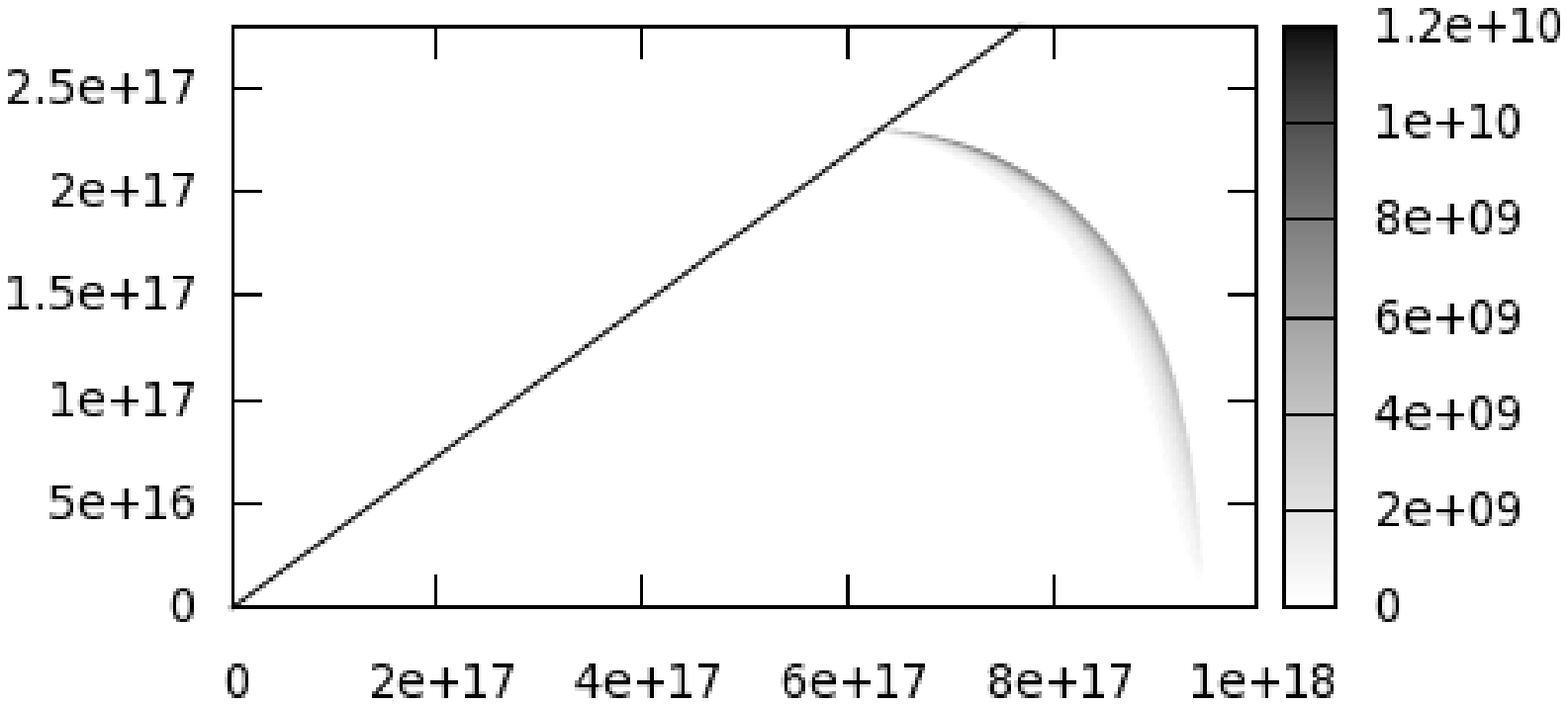}
\includegraphics[width=1.0\columnwidth]{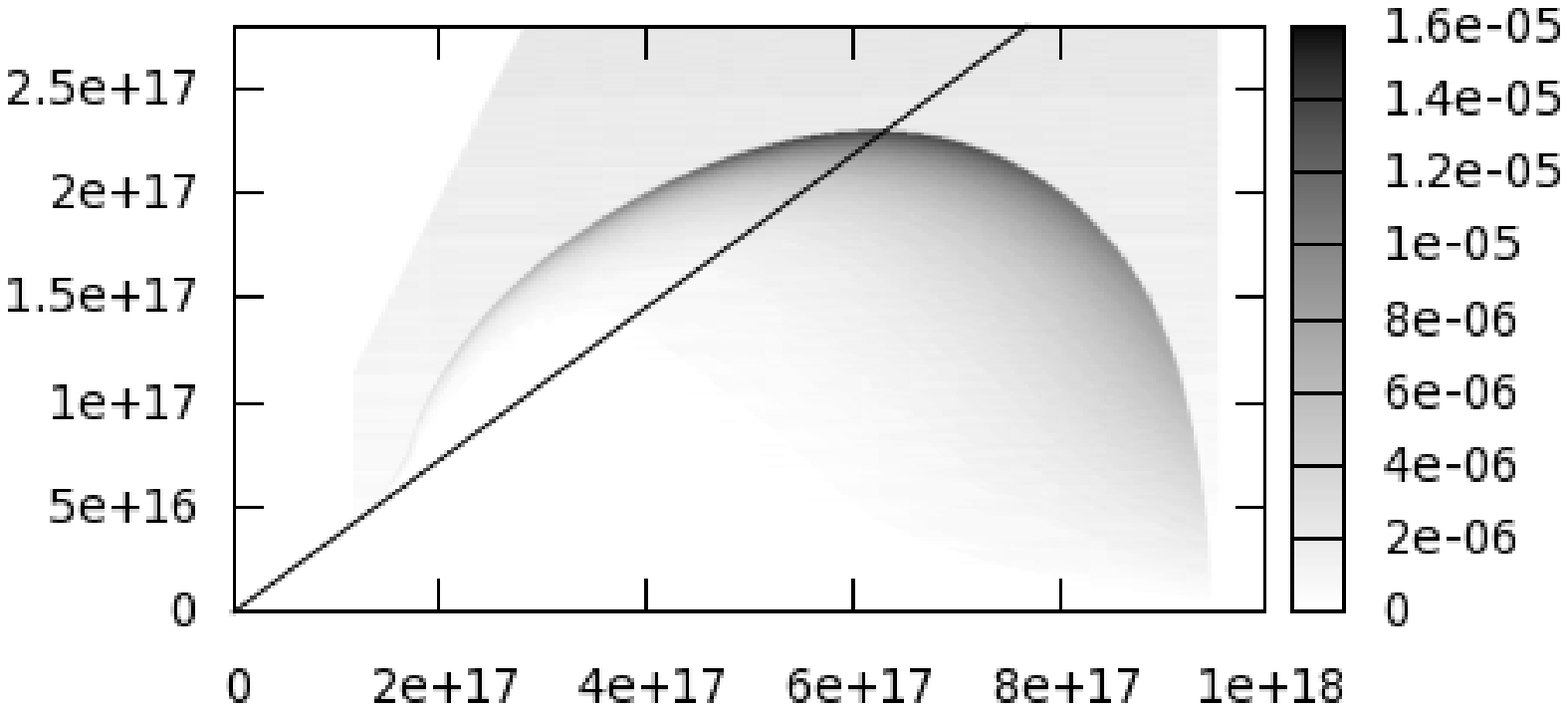}
\caption{Region contributing to the observed radio flux in two dimensions. The top plot shows the angle integrated, absorption corrected emissivity coefficients for $1.4 \cdot 10^9$ Hz at 28 days in observer time, similar to the plots in fig. \ref{emission_coefficients_figure}. The lower plot shows the angle integrated lab frame density. The diagonal line in both plots indicates a half opening angle of 20 degrees.}
\label{emission_2D_figure}
\end{figure}

In figure \ref{emission_2D_figure} we explore the radio emission from the 2D simulation in some more detail. The top figure shows that the contributing region at this strongly self-absorbed frequency actually stays within the cone of half opening angle 20 degrees. It would however be wrong to conclude from this that almost no lateral spreading has occured. In the lower figure we plot the ring integrated observer frame density, and this clearly reveals that the jet has spread out noticeably already. However, as explained in \cite{Zhang2009}, it is the mildly relativistic jet material behind the shock that undergoes more sideways expansion and this is not the material that contributes the most to the observed flux. Fig. \ref{emission_2D_figure} confirms the effect of the optical depth, which we argued to be chiefly responsible for the delay in observed jet break time. Since material has moved sideways, the optical depth through the fluid within the cone decreases and the delay in jet break time due to high optical depth becomes less strong.

\section{Summary and conclusions}
\label{jetbreaks_summary_section}
We have presented simulation results for high-resolution adaptive mesh simulations of afterglow blast waves in one and two dimensions. We have used explosion parameters that nominally apply to GRB 030329, but these were derived by earlier authors using a purely analytical model (in \cite{vanEerten2009c} we have shown that the difference between analytical model and simulation is significant. A paper improving fit models using simulation results is currently in preparation). The aim of these simulations is to study the achromaticity of the afterglow jet break. We have found that the jet break is chromatic between frequencies below both the self-absorption break $\nu_{a}$ and the synchrotron break $\nu_m$ and above these breaks.

A result that is of direct practical use is an improvement on the relationship between observed jet break time and opening angle. We show how the conventional analysis for jet breaks in optically thin jets, systematically underestimate the jet opening angle. For a homogeneous medium, this amounts to a factor 1.3, but we give an expression for general circumburst medium density structure.

The difference in jet break time between optically thick and optically thin frequencies is explored in some detail. We have calculated emission profiles for the jet at various observer frequencies at ten days observer time. From these calculations it is shown that the region of the blast wave dominating the observed flux moves around for different frequencies. At high frequencies, this region lies more to the edge of the jet and to early emission times, whereas at high optical depth and frequencies below the synchrotron break the contributing region moves to the front and center of the blast wave -and therefore remains within the jet cone long after the high frequency region falls outside the jet cone.

The slope of the jet break will be different for different observer frequencies. If limb brightening is strong, the jet break slope for hard-edged jets (without lateral expansion) will initially overshoot its asymptotic value and then slowly evolve to its asysmptotic limit, which lies $(3-k)/(4-k)$ below the corresponding slope for the spherical case at the same observer time.

We have confirmed our results in two dimensions. The jet break in the radio is still postponed, but less so than for hard-edged jets. At 28 days in observer time, the effectively contributing region remains within the initial opening angle, even though lateral spreading of the fluid has already progressed noticably. As a result of the lateral spreading, the optical depth through the fluid decreases.

The chromaticity of the jet breaks is a result of the detailed interplay between the synchrotron radiation mechanisms and the fluid dynamics. Models that treat synchrotron radiation in a simplified manner, like \cite{Zhang2009}, which does not include self-absorption and does not locally calculate the cooling times, will not reveal a chromatic jet break. \cite{Granot2007} does not discuss self-absorption in the context of the jet break.

Finally we note that we have in this entire paper assumed the observer to lie precisely on the jet axis. Although the observer angle will be small in practice compared to the jet opening angle (in order to be able to observe early afterglow and prompt emission in the first place), this is nevertheless likely to have a profound effect on the shape and timing of the jet break. We are expanding the radiation code to deal with off-axis observers and will report the results in our follow-up study.

\section{Acknowledgements}
HJvE and RAMJW thank Konstantinos Leventis for useful discussion. This research was supported by NWO Vici grant 639.043.302 (RAMJW, KL), NOVA project 10.3.2.02 (HJvE). HJvE was also supported in part by NASA under Grant No. 09-ATP09-0190 issued through the Astrophysics Theory Program (ATP). ZM performed computations on the JADE (CINES) on DARI project and on the  KU Leuven High Performance computing cluster VIC, and acknowledges financial support from the FWO, grant G.0277.08, from the GOA/2009/009 and HPC Europa.

\appendix
\section{Numerical approach and resolution}
\label{numerical_approach_section}

We have used the same basic approaches to the dynamics and to the calculation of the synchrotron radiation as described in \citet{vanEerten2009c}. However, upon further experimenting we have found that a number of modifications of electron cooling improve the accuracy of our results. We still follow the local evolution of the upper cut-off Lorentz factor $\gamma'_M$ of the shock-accelerated particle distribution with an advection equation, but now use
\begin{equation}
\frac{\partial}{\partial t} \left( \frac{\gamma (\rho')^{4/3}}{\gamma'_M} \right) + \frac{\partial}{\partial x^i} \left( \frac{ \gamma (\rho')^{4/3} v^i}{\gamma'_M} \right) = \alpha \cdot (\rho' )^{4/3} \cdot (B')^2,
\end{equation}
instead of equation (A6) from \citet{vanEerten2009c}. This equation can be derived as usual from combining the continuity equation and the kinetic equation. Also we now no longer explicitly inject hot electrons at the shock front during the simulation, but do this implicitly via the initial conditions of the simulation. We do this by setting $\gamma'_M$ initially equal to $10^{10}$ everywhere outside the shock. Because the unshocked material is very cold, and the magnetic field strength is linked to the thermal energy density via $\epsilon_B$, synchrotron cooling will not change $\gamma'_M$ outside of the shock. Once a shock passes, the fluid is heated and electron cooling automatically sets in directly. If we now ignore unshocked parts of the fluid grid (i.e. cold, nonmoving areas that are resolved with only a few refinement levels) when calculating emission, we have an algorithm to calculate synchrotron radiation including electron cooling, but where we do not need to worry about seeking out the shock front during each iteration of the RHD simulation. With these alterations, our method has moved closer to that implemented by \citet{Downes2002} 

\begin{figure}
\centering
\includegraphics[width=1.0\columnwidth]{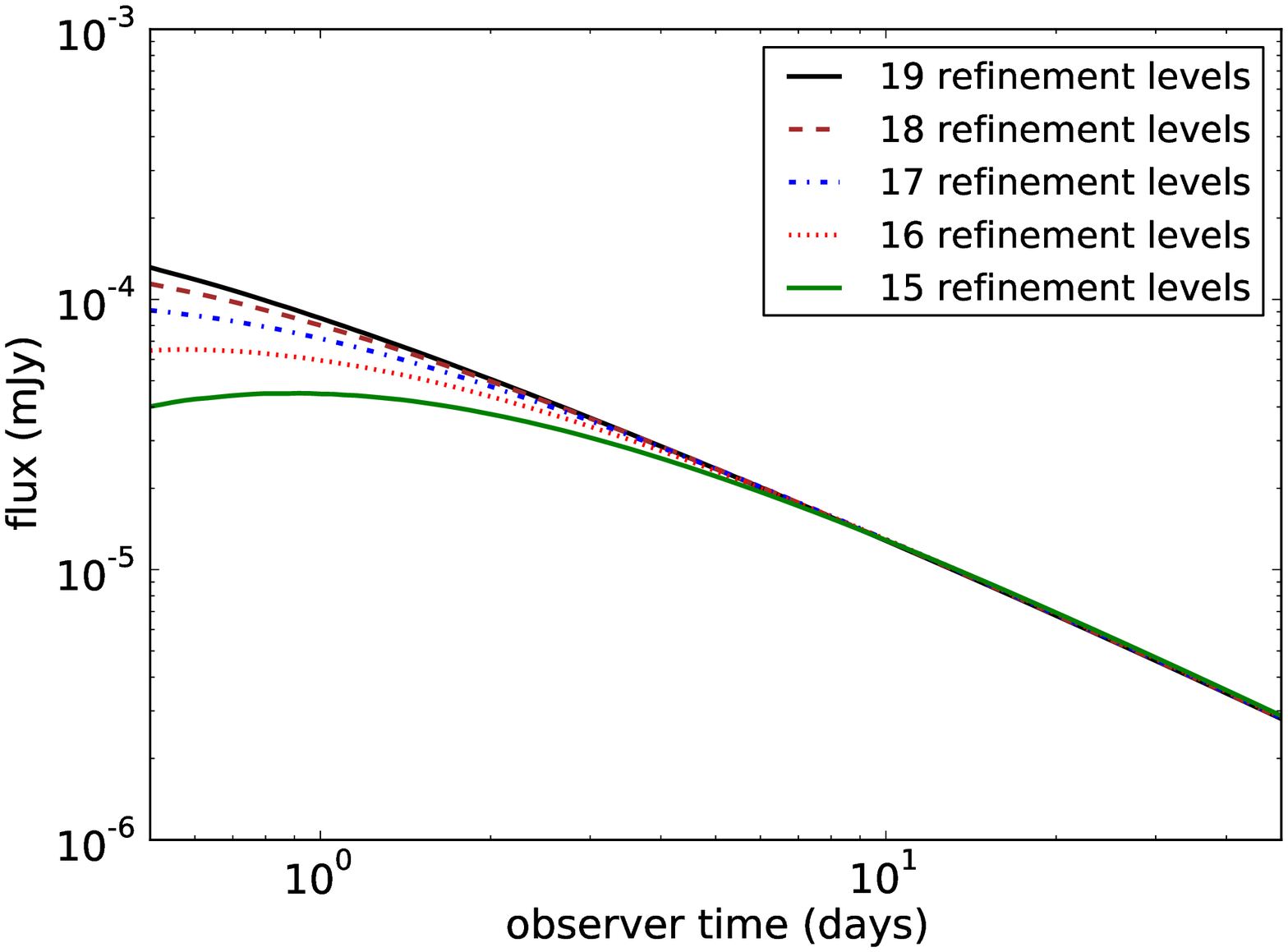}
\caption{Light curves calculated at different refinement levels, for a spherical explosion. We have kept the refinement levels of the fluid simulation and the radiation calculation identical. The observer frequency has been set at $5 \cdot 10^{17}$ Hz, well above the cooling break.}
\label{res_lightcurve_figure}
\end{figure}

\begin{figure}
\centering
\includegraphics[width=1.0\columnwidth]{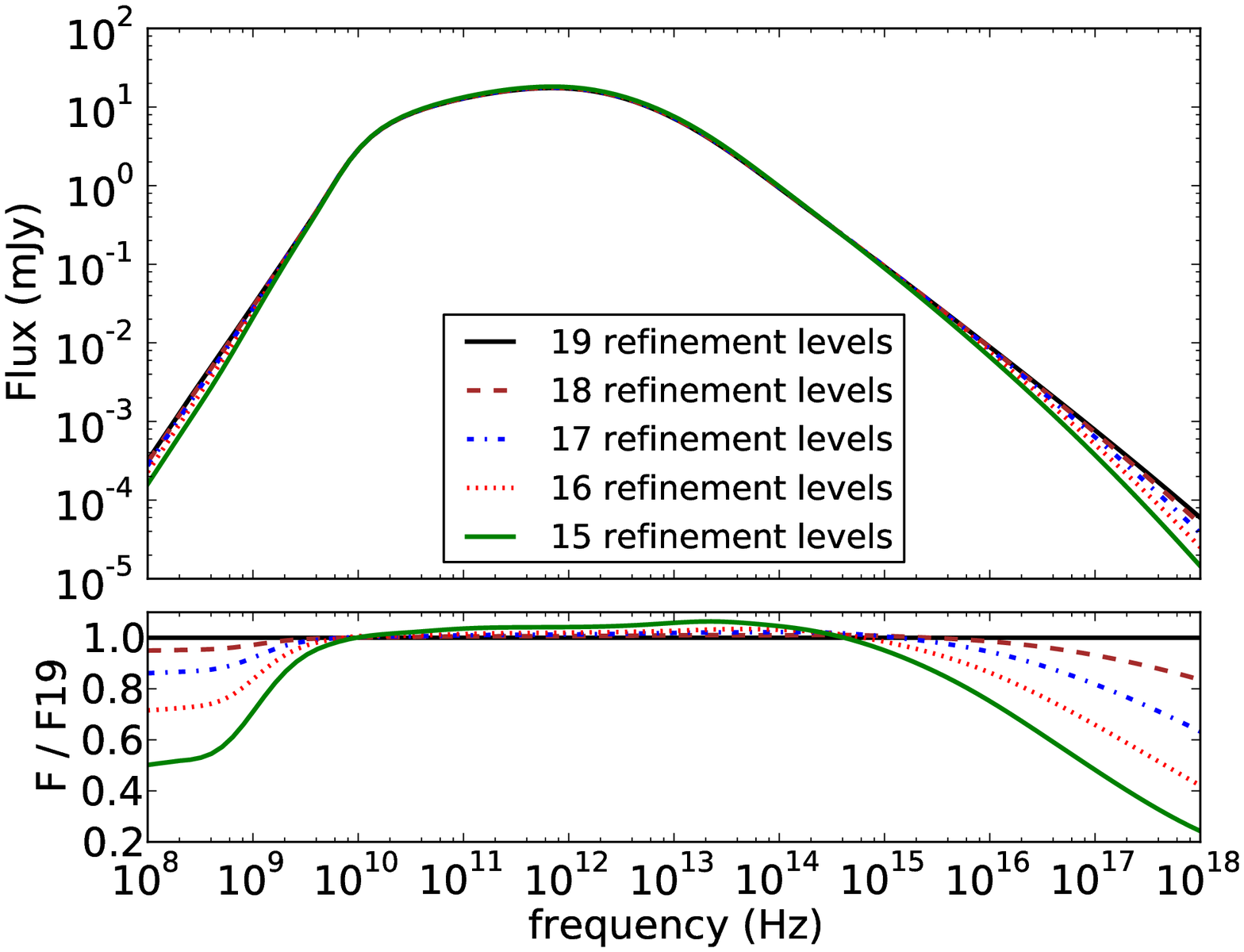}
\caption{Spectra calculated at different refinement levels, for a spherical explosion. The observer time has been set at 0.5 days. The lower plot shows the monochromatic flux for a given refinement level as fraction of the flux from the simulation at 19 refinement levels. As with the light curve, the refinement levels of the fluid simulation and the radiation calculation have been kept identical.}
\label{res_spectrum05_figure}
\end{figure}

To show the numerical validity of our results and check the resolution, we have performed calculations at different refinement levels. In figure \ref{res_lightcurve_figure} we show light curves at observer frequency $5 \cdot 10^{17}$ Hz for a spherical explosion. We show this high frequency because the hot region that dictates the spectrum above the cooling break is the hardest to resolve. This is illustrated by figure \ref{res_spectrum05_figure}, which shows the spectrum for a spherical explosion at observer time 0.5 days, the earliest time used in plots in this paper. Because the blast wave width is smaller at earlier times, this is therefore also where any resolution issues should be most apparent. The light curve in fig. \ref{res_lightcurve_figure} shows that the simulations quickly converge for the different refinement levels at later times. When the jet breaks occur, around a few days or so depending on the chosen jet opening angle, the convergence of the light curves is sufficient to show that the results of this paper remain unaltered under further increase in resolution. We also note that convergence is achieved at an earlier time for frequencies below the cooling break, as can be seen from the spectrum, which confirms that electron cooling and fluid evolution occur on different spatial and temporal scales (as one would theoretically expect).

We have also tested the temporal resolution of the simulations by comparing light curves from a datasets with 1000 snapshots to light curves from a dataset with 10,000 snapshots. For 1000 snapshots the temporal resolution is $3.7 \cdot 10^4$ s. and for 10,000 it is $3.7 \cdot 10^3$ s. in emission time. The resolution in observer time is better than the resolution in emission time, due to angular smearing and compression of the signal. In practice the resulting flux between 1000 and 10,000 turns out to differ less than one percent at early times (0.5 days). This difference only becomes smaller at later observer times.

\bibliographystyle{mn2e}
\bibliography{hveerten}

\end{document}